\documentclass[aps,prc,onecolumn,nofootinbib,superscriptaddress,floatfix]{revtex4-2}

\usepackage{graphicx}
\usepackage{amsmath}
\usepackage{amssymb}
\usepackage{bm}
\usepackage{hyperref}
\usepackage{placeins}
\hypersetup{colorlinks=true,linkcolor=blue,citecolor=blue,urlcolor=blue}

\begin{document}

\title{Quark--hadron duality in inclusive electron--proton scattering
at high $Q^{2}$: structure functions and truncated moments from CLAS12}

\author{Y.~Wunderlich}
\affiliation{University of Connecticut, Storrs, Connecticut 06269, USA}
\author{A.~Bulgakov}
\affiliation{University of Connecticut, Storrs, Connecticut 06269, USA}
%\author{V.~Klimenko}
%\affiliation{University of Connecticut, Storrs, Connecticut 06269, USA}
\author{K.~Joo}
\affiliation{University of Connecticut, Storrs, Connecticut 06269, USA}
\author{T.-S.~H.~Lee}
\affiliation{Physics Division, Argonne National Laboratory,
Argonne, Illinois 60439, USA}
\author{V.~I.~Mokeev}
\affiliation{Thomas Jefferson National Accelerator Facility,
Newport News, Virginia 23606, USA}

% additional authors here

\date{\today}

\begin{abstract}
We present a high-precision study of quark--hadron duality in inclusive electron--proton
scattering in the nucleon resonance region, extending to
$Q^{2}\approx10~\mathrm{GeV}^{2}$, based on recent CLAS12
cross-section measurements at Jefferson Lab. The data, taken with a
$10.6$~GeV electron beam, span nine bins in four-momentum transfer over
$2.55 \le Q^{2} \le 10.4~\mathrm{GeV}^{2}$ and cover the full resonance
region from the pion-production threshold up to
$W\approx2.5~\mathrm{GeV}$. To extend the hadronic description into the
CLAS12 kinematic regime, we develop a phenomenological high-$Q^{2}$
extension of the Argonne National Laboratory--Osaka University
(ANL-Osaka) dynamical coupled-channels framework, constrained by the
measured cross sections and anchored to the original ANL-Osaka
calculation at $Q_{0}^{2}=2.774~\mathrm{GeV}^{2}$. The extended
framework enables a phenomenologically constrained
longitudinal--transverse decomposition and extraction of the proton
structure function $F_{2}(W,Q^{2})$, from which we evaluate
$W$-truncated Cornwall--Norton moments $M_{2}(Q^{2})$ for the first,
second, and third resonance regions as well as the high-$W$ region.
Comparison with the CJ15 global QCD analysis, including target-mass and
higher-twist corrections, demonstrates overall consistency at the
cross-section, structure-function, and truncated-moment levels across
the kinematic range probed, providing quantitative evidence for both local and global
quark--hadron duality and extending previous duality tests to
substantially higher $Q^{2}$. 
A further result of this analysis is the identification of a
threshold effect in the partonic calculation: the finite-$Q^{2}$
corrections do not enforce the physical pion-production threshold, and
the residual discrepancy in the first resonance region is consistent
with this effect rather than a breakdown of duality, indicating that
threshold-constrained global analyses could sharpen future duality
tests. Within the coupled-channel description, the single-pion channel
alone underestimates the inclusive resonance-region strength above the
$\Delta(1232)$, which is carried predominantly by the multi-meson
channels, as required for duality.
\end{abstract}

\maketitle

\raggedbottom

% =====================================================================
\section{Introduction}
\label{sec:intro}
% =====================================================================

Inclusive electron--proton scattering has long provided a primary avenue
for investigating the internal structure of the nucleon across a wide
range of distance scales. In the deep-inelastic scattering (DIS) regime,
the cross sections are successfully described in terms of parton
distribution functions (PDFs) governed by perturbative quantum
chromodynamics (QCD). At lower invariant masses, $W \lesssim
2~\mathrm{GeV}$, the cross sections exhibit pronounced resonance
structures associated with the excitation of nucleon resonances,
reflecting the strongly coupled, nonperturbative dynamics of QCD.
A remarkable empirical connection between these two regimes is provided
by \emph{quark--hadron duality}, first observed by Bloom and
Gilman~\cite{BloomGilman1970,BloomGilman1971}. In inclusive scattering,
this phenomenon manifests itself through the observation that structure
functions measured in the resonance region, when appropriately averaged
over the invariant mass $W$, reproduce the same scaling behavior as
those measured in the DIS region. Subsequent experimental measurements
and theoretical investigations have shown that duality holds not only in
a global sense---when averaged over broad resonance regions---but also,
under suitable conditions, locally within individual resonance
regions~\cite{MelnitchoukEntKeppel,Malace2009}. From a theoretical
perspective, quark--hadron duality can be understood within the
framework of the operator product expansion (OPE), where higher-twist
contributions become progressively suppressed with increasing
four-momentum transfer $Q^{2}$, allowing the resonance dynamics to
reproduce leading-twist behavior upon
averaging~\cite{MelnitchoukEntKeppel}.

Quantitative tests of quark--hadron duality require precise measurements
of inclusive observables in the resonance region over a broad range of
$Q^{2}$, together with theoretical descriptions capable of connecting
the hadronic and partonic pictures. On the partonic side, modern global
QCD analyses of PDFs have demonstrated that leading-twist calculations,
supplemented by target-mass corrections (TMC) and phenomenological
higher-twist (HT) terms, can describe inclusive DIS data down to
relatively low values of $Q^{2}$ and invariant mass~\cite{HillerBlin:2021hhm,Accardi2016,GeorgiPolitzer1976,Schienbein2008}. This, in turn,
makes it possible to evaluate structure functions and their truncated
moments within a partonic framework and to compare them directly with
resonance-region measurements.

Recent measurements of inclusive electron--proton scattering performed
with the CLAS12 detector~\cite{CLAS12} at Jefferson Lab have
substantially extended the available data in the nucleon resonance
region. These measurements, carried out by V.~Klimenko
\textit{et al.}~\cite{KlimenkoCLAS12} using CLAS12 Run~Group~A (RGA) data taken
at a beam energy $E = 10.6~\mathrm{GeV}$, provide absolute cross
sections over the invariant-mass range from the pion-production
threshold up to $W \approx 2.5~\mathrm{GeV}$ for nine bins in
four-momentum transfer spanning $2.55 \le Q^{2} \le
10.4~\mathrm{GeV}^{2}$. Owing to the large acceptance of CLAS12, the
data cover the full resonance region at each value of $Q^{2}$, offering
a unique opportunity to investigate the evolution of nucleon structure
in the transition region between the hadronic and partonic descriptions
and to test duality at momentum transfers well beyond those of earlier
dedicated measurements in Hall~C~\cite{Malace2009,Tvaskis2018} and with
CLAS~\cite{Osipenko2003}. These measurements allow a direct test of whether the consistency
between hadronic and partonic descriptions observed in earlier duality
studies persists at substantially higher momentum
transfers.

The primary goal of quark--hadron duality studies is to assess whether
descriptions of nucleon structure based on hadronic degrees of freedom
and those based on partonic degrees of freedom yield consistent results
for inclusive observables. Addressing this question requires a framework
in which the same physical quantities can be evaluated within both
pictures and compared on an equal footing. On the hadronic side, this
necessitates a theoretical description that explicitly incorporates
nucleon resonances and is capable of reproducing the measured inclusive
cross sections in the resonance region. Such a description can then be
used to compute structure functions and their truncated moments, which
represent averaged resonance contributions over specified
invariant-mass regions. On the partonic side, the same observables can
be evaluated using global QCD analyses of PDFs, including finite-$Q^{2}$
corrections.

The hadronic description adopted in this work is provided by the
Argonne National Laboratory--Osaka University (ANL-Osaka) dynamical
coupled-channels (DCC)
model~\cite{SatoLee1996,Matsuyama2007,JuliaDiaz2007,Kamano2009a,Suzuki2010,Kamano2013,Nakamura2015,Kamano2016}; the abbreviation AO is used for this model in equations throughout this
paper. The ANL-Osaka model builds on a five-channel EBAC framework for
$\pi N$ and $\gamma N$ reactions developed by an international
collaboration within the Excited Baryon Analysis Center (EBAC) program
at Jefferson Laboratory, and was later extended to an eight-channel model
that also describes $\nu N$ and $K N$ reactions. In this framework the
inclusive cross section is constructed from the $\pi N$, $\eta N$,
$K\Lambda$, $K\Sigma$, and $\pi\pi N$ channels, the last of which
includes resonant $\pi\Delta(1232)$, $\rho N$, and $\sigma N$
components, with parameters constrained by global fits to $\pi N$ and
$\gamma N$ data. The original ANL-Osaka description of inclusive
electron--proton scattering is, however, limited to
$Q^{2} \lesssim 3~\mathrm{GeV}^{2}$, because the ANL-Osaka
parametrization of the resonance electromagnetic form factors
$F_{N^{*},\gamma N}(Q^{2})$ was established only up to this range in
the original analysis of exclusive $p(e,e'\pi)N$
data~\cite{Nakamura2015}.

A central element of the present analysis is therefore a
phenomenological extension of the ANL-Osaka model to high $Q^{2}$,
anchored to the ANL-Osaka calculation at
$Q_{0}^{2} = 2.774~\mathrm{GeV}^{2}$ and constrained directly by the
CLAS12 cross sections at higher $Q^{2}$. This extension is empirical: it
provides a controlled high-$Q^{2}$ interpolation anchored to the
ANL-Osaka calculation, rather than a new dynamical coupled-channels
solution. Using this extended
description, we perform a phenomenologically constrained longitudinal--transverse
(L/T) decomposition, extract the proton structure function
$F_{2}(W,Q^{2})$, and compute $W$-truncated Cornwall--Norton moments
$M_{2}(Q^{2})$ for four $W$ ranges corresponding to the first, second,
and third resonance regions and the high-$W$ region beyond. The results are compared with two variants of the CJ15 next-to-leading-order (NLO) global
QCD analysis~\cite{Accardi2016}, evaluated using LHAPDF~6.5.4~\cite{CJ15}:
the NLO leading-twist baseline (denoted CJ15NLO+LT) and the same calculation
supplemented by target-mass and higher-twist corrections (CJ15NLO+TMC+HT).

This is the first quantitative study of quark--hadron duality that combines
recent CLAS12 measurements covering the full resonance region and
extending to $Q^{2}\approx10~\mathrm{GeV}^{2}$ with a coupled-channels
hadronic description that incorporates both single-meson and
multi-meson final states. The analysis establishes consistency between hadronic and
partonic descriptions simultaneously at the level of inclusive cross
sections, extracted structure functions, and truncated moments. In
addition, the single-pion channel of the coupled-channel solution alone
underestimates the inclusive resonance-region strength above the
$\Delta(1232)$, demonstrating that the multi-meson channels are
essential for reproducing the strength required for local quark--hadron
duality. These advances extend previous duality
studies to substantially higher momentum transfers and provide new
quantitative evidence for both local and global quark--hadron duality.

This paper is organized as follows. In Sec.~\ref{sec:AO_native} we
summarize the ANL-Osaka dynamical coupled-channels model and its
performance in describing the CLAS12 inclusive cross section at
$Q^{2} = 2.774~\mathrm{GeV}^{2}$. Section~\ref{sec:AO_extension}
describes the phenomenological extension of the ANL-Osaka model to high
$Q^{2}$ and the fitting procedure used to extract the structure
function $F_{2}$. The extended description is confronted with the
measured inclusive cross sections at multiple $Q^{2}$ bins in
Sec.~\ref{sec:xsec_4bins}. The extracted $F_{2}(W,Q^{2})$ and the
resulting picture of local quark--hadron duality are presented in
Sec.~\ref{sec:F2}, and the $W$-truncated moments $M_{2}(Q^{2})$ are
discussed in Sec.~\ref{sec:moments}. A summary and outlook are given in
Sec.~\ref{sec:summary}.

% =====================================================================
\section{The ANL-Osaka model and its performance at
$Q^{2} \le 2.774~\mathrm{GeV}^{2}$}
\label{sec:AO_native}
% =====================================================================

The ANL-Osaka dynamical coupled-channels (DCC) framework provides a
unified description of meson--baryon reactions and electromagnetic
processes in the nucleon resonance region. Its electroproduction
amplitudes have a factorized form,
\begin{equation}
T_{MB,\gamma^{*}N}(W,Q^{2})
\;=\;
t^{R}_{MB,\gamma^{*}N}(W,Q^{2})
\;+\;
t^{NR}_{MB,\gamma^{*}N}(W,Q^{2}),
\label{eq:T_AO}
\end{equation}
where the resonant amplitude is
\begin{equation}
t^{R}_{MB,\gamma^{*}N}(W,Q^{2})
\;=\;
\left[
\sum_{N^{*}}
\frac{\bar{\Gamma}^{\dagger}_{N^{*},MB}(W)\,
\bar{\Gamma}_{N^{*},\gamma N}(W)}
{W - M^{0}_{N^{*}} - \Sigma_{N^{*}}(W)}
\right] F_{N^{*},\gamma N}(Q^{2}),
\label{eq:t_R}
\end{equation}
and the nonresonant amplitude has an analogous structure with form
factors $F^{n}_{MB,\gamma N}(Q^{2})$ associated with each Born-type
contribution. In Eq.~(\ref{eq:t_R}), $\bar{\Gamma}_{N^{*},ab}(W)$ is
the dressed vertex for the $N^{*} \to ab$ decay, $M^{0}_{N^{*}}$ is the
bare resonance mass, $\Sigma_{N^{*}}(W)$ is the mass shift generated by
the nonresonant interactions, and $F_{N^{*},\gamma N}(Q^{2})$ is the
$\gamma N \to N^{*}$ electromagnetic transition form factor. Within the
original ANL-Osaka analysis, the $\gamma N \to N^{*}$ form factors are
parametrized as a polynomial in $Q^{2}$,
\begin{equation}
F_{N^{*},\gamma N}(Q^{2})
\;=\;
\sum_{n=0}^{n_{\max}} c^{N}_{n}\,(Q^{2})^{n},
\label{eq:FN_poly}
\end{equation}
with the normalization at the photon point,
$F_{N^{*},\gamma N}(Q^{2}=0) = c^{N}_{0}$, fixed by global fits to the
photo- and pion-electroproduction world
data~\cite{Kamano2013,Nakamura2015,Kamano2016}. Similar
coupled-channel descriptions have been developed by the Bonn--Gatchina
group~\cite{Anisovich2012} and the J\"ulich--Bonn--Washington
group~\cite{Roenchen2013,Mai2021}; in the present work the ANL-Osaka
model is adopted as the hadronic-basis input.

\begin{figure}[htbp]
\centering
\includegraphics[width=0.85\columnwidth]{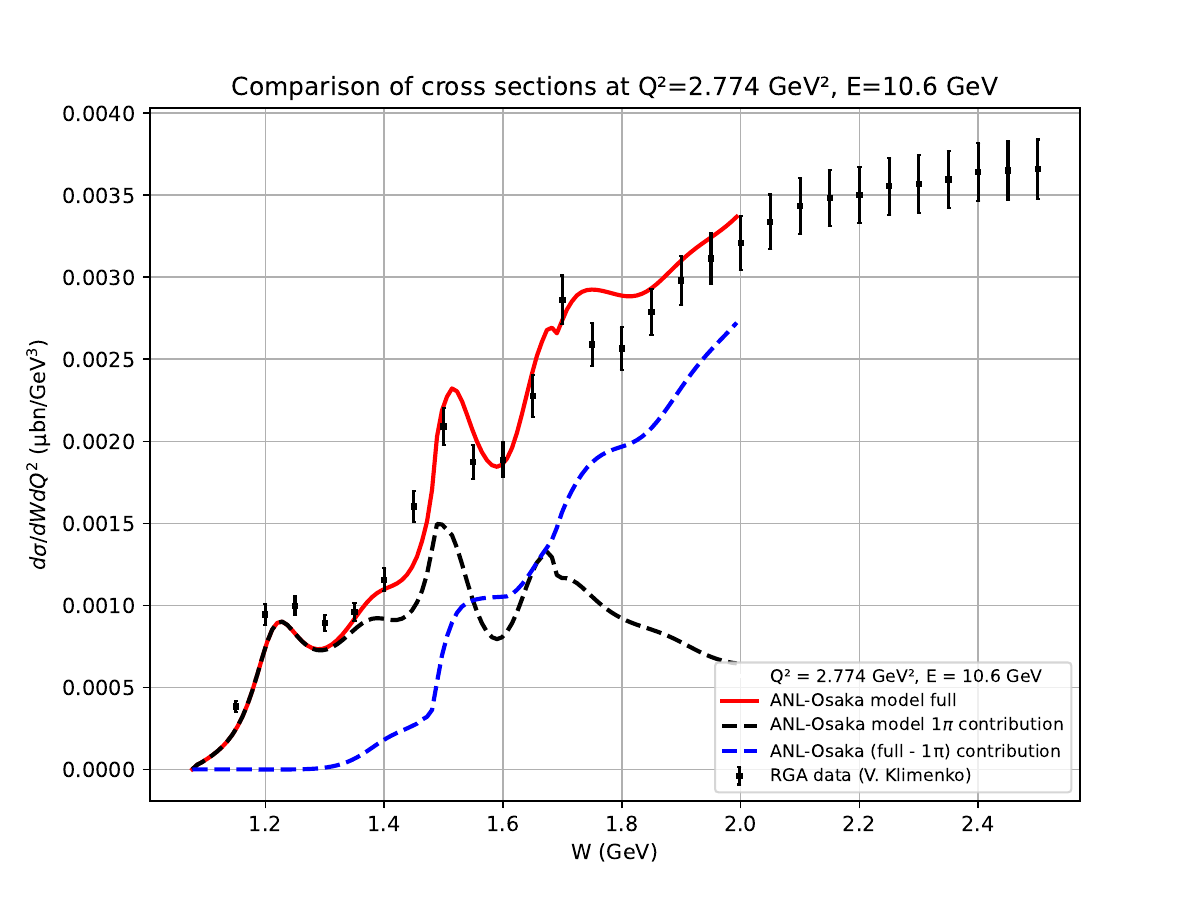}
\caption{Inclusive cross section $d^{2}\sigma/dW\,dQ^{2}$ at
$Q^{2} = 2.774~\mathrm{GeV}^{2}$ and $E = 10.6~\mathrm{GeV}$.
Black points: CLAS12 RGA data~\cite{KlimenkoCLAS12}. Red solid:
full ANL-Osaka model. Black dashed: $1\pi$ contribution. Blue
dashed: sum of the multi-meson channels (two-pion, kaon, and
heavier final states). The full ANL-Osaka calculation reproduces
the resonance-region data within uncertainties; the channel decomposition shows that multi-meson channels provide the dominant contribution to the inclusive strength above $W\approx1.5~\mathrm{GeV}$.}
\label{fig:xsec_2p77_decomposition}
\end{figure}

Figure~\ref{fig:xsec_2p77_decomposition} shows the CLAS12 inclusive
cross section $d^{2}\sigma/dW\,dQ^{2}$ at
$Q^{2} = 2.774~\mathrm{GeV}^{2}$ and $E = 10.6~\mathrm{GeV}$, compared
with the full ANL-Osaka calculation (red solid). The model reproduces
the position and magnitude of the $\Delta(1232)$, second-resonance, and
third-resonance peaks observed in the data within the experimental
uncertainties. The channel decomposition reveals an important physical
feature: the single-pion ($1\pi$) subchannel (black dashed) dominates
the cross section at low $W$ but is rapidly overtaken by the
multi-meson channels (blue dashed), which together carry the bulk of
the strength above $W \approx 1.5~\mathrm{GeV}$. The exclusive $1\pi$ channel alone underestimates the inclusive cross
section by roughly a factor of two near
$W \approx 1.7~\mathrm{GeV}$, indicating that high-mass $N^{*}$ states
with sizable $\pi\Delta$, $\rho N$, and $\sigma N$ branching fractions
must be retained for a quantitative description of the inclusive
strength above the $\Delta(1232)$. This comparison demonstrates explicitly that multi-meson channels are
essential for reproducing the inclusive resonance-region strength and
for realizing quark--hadron duality above the $\Delta(1232)$ region.
The full ANL-Osaka calculation therefore provides the natural
hadronic-basis input for the high-$Q^{2}$ extension developed in the
following section.

% =====================================================================
\section{Phenomenological extension of the ANL-Osaka model to high
$Q^{2}$}
\label{sec:AO_extension}
% =====================================================================

The ANL-Osaka model is extended to the CLAS12 kinematics in two stages,
both anchored at $Q_{0}^{2} = 2.774~\mathrm{GeV}^{2}$ and evolved to
higher $Q^{2}$ through the same exponential-scaling Breidenbach--Kuti
Ansatz in $\omega'$. In Fit~1 (Sec.~\ref{sec:fit1}) the inclusive cross
section is decomposed into single-pion and multi-pion contributions,
with their ratio $R_{1\pi/2\pi}$ fixed to the ANL-Osaka value at
$Q_{0}^{2}$, over the resonance region $W \lesssim 2.0~\mathrm{GeV}$
where the ANL-Osaka model is defined. In Fit~2 (Sec.~\ref{sec:fit2}) the
inclusive cross section is decomposed instead into longitudinal and
transverse virtual-photon components, with the ratio $R_{LT}$ softly
constrained to a reference value at $Q_{0}^{2}$, and the description is
extended to $W \le 2.5~\mathrm{GeV}$. The two fits decompose the same
inclusive cross section along different axes: Fit~1 resolves its
single-pion/multi-pion content, whereas Fit~2 resolves the longitudinal
and transverse components of the total cross section but not its
single-pion/multi-pion content. The longitudinal--transverse separation
of Fit~2 yields the proton structure function $F_{2}(W,Q^{2})$, from
which the $W$-truncated moments that enter the duality analysis are
computed.

Since the resonance electromagnetic form factors of
Eq.~(\ref{eq:FN_poly}) have been determined only for $Q^{2} \le
3~\mathrm{GeV}^{2}$, a direct application of the ANL-Osaka model at the
higher $Q^{2}$ values covered by CLAS12 is not currently possible.
Following the strategy of earlier SLAC analyses of inclusive
$p(e,e')$ scattering~\cite{BreidenbachKuti1972}, we exploit the
factorized form of Eqs.~(\ref{eq:T_AO})--(\ref{eq:t_R}), in which each
electroproduction amplitude separates into a $W$-dependent hadronic
part and a $Q^{2}$-dependent form factor. This motivates the
approximate representation of the inclusive structure functions,
\begin{equation}
W_{i}(Q^{2},W)
\;\simeq\;
\bigl|F^{\rm AO}(W,Q^{2})\bigr|^{2}
\sum_{MB}
\bigl|
t^{R}_{MB,\gamma N}(W,Q^{2}=0)
+ t^{NR}_{MB,\gamma N}(W,Q^{2}=0)
\bigr|^{2},
\label{eq:W_factorized}
\end{equation}
in which the $W$ dependence is fixed by the photoproduction amplitudes
generated from the ANL-Osaka model, and the averaged $Q^{2}$ evolution
is encoded in the effective form factor $|F^{\rm AO}(W,Q^{2})|^{2}$.

\subsection{Fit~1: single-pion versus multi-pion decomposition
($R_{1\pi/2\pi}$ fixed at $Q_{0}^{2}$)}
\label{sec:fit1}

In the first stage we parametrize the inclusive cross section as a sum
of evolved single-pion and multi-pion contributions, each anchored at
$Q_{0}^{2}$,
\begin{equation}
\sigma(W,Q^{2})
\;=\;
\tilde\sigma^{1\pi}(W,Q^{2})
+ \tilde\sigma^{2\pi}(W,Q^{2}),
\label{eq:fit1}
\end{equation}
with the evolved single-pion and multi-pion contributions
\begin{align}
\tilde\sigma^{1\pi}(W,Q^{2})
&= e^{-\alpha(Q^{2}-Q_{0}^{2})}\,
\frac{F_{2}\bigl(\omega'(Q^{2},W);c_{1},c_{2},c_{3}\bigr)}
     {F_{2}\bigl(\omega'(Q_{0}^{2},W);c_{1},c_{2},c_{3}\bigr)}
\,\sigma^{1\pi}(W,Q_{0}^{2}),
\label{eq:fit1_1pi}\\[4pt]
\tilde\sigma^{2\pi}(W,Q^{2})
&= e^{-\beta(Q^{2}-Q_{0}^{2})}\,
\frac{F_{2}\bigl(\omega'(Q^{2},W);c'_{1},c'_{2},c'_{3}\bigr)}
     {F_{2}\bigl(\omega'(Q_{0}^{2},W);c'_{1},c'_{2},c'_{3}\bigr)}
\,\sigma^{2\pi}(W,Q_{0}^{2}),
\label{eq:fit1_2pi}
\end{align}
where $\sigma^{1\pi}(W,Q_{0}^{2})$ and $\sigma^{2\pi}(W,Q_{0}^{2})$ are
the ANL-Osaka single-pion and multi-pion cross sections at the anchor
point, $\omega'(Q^{2},W) = 1 + W^{2}/Q^{2}$, and the modified
Breidenbach--Kuti function is~\cite{BreidenbachKuti1972}
\begin{equation}
F_{2}\bigl(\omega';c_{1},c_{2},c_{3}\bigr)
\;=\;
\sum_{i=1}^{3} c_{i}
\left(1 - \frac{1}{\omega'}\right)^{2+i}.
\label{eq:F2_BK_3}
\end{equation}
By construction, this parametrization reproduces the ANL-Osaka model at
$Q^{2} = Q_{0}^{2}$, while the exponential factors
$e^{-\alpha(Q^{2}-Q_{0}^{2})}$ and $e^{-\beta(Q^{2}-Q_{0}^{2})}$
stabilize the extrapolation by suppressing the otherwise divergent
behavior of the $\omega'$ parametrization at large $Q^{2}$. This
parametrization yields good fits to the CLAS12 inclusive cross sections
across the full $Q^{2}$ range, with a data-only
$\chi^{2}/\mathrm{ndf} \approx 2$.

\begin{figure}[htbp]
\centering
\includegraphics[width=0.95\textwidth]{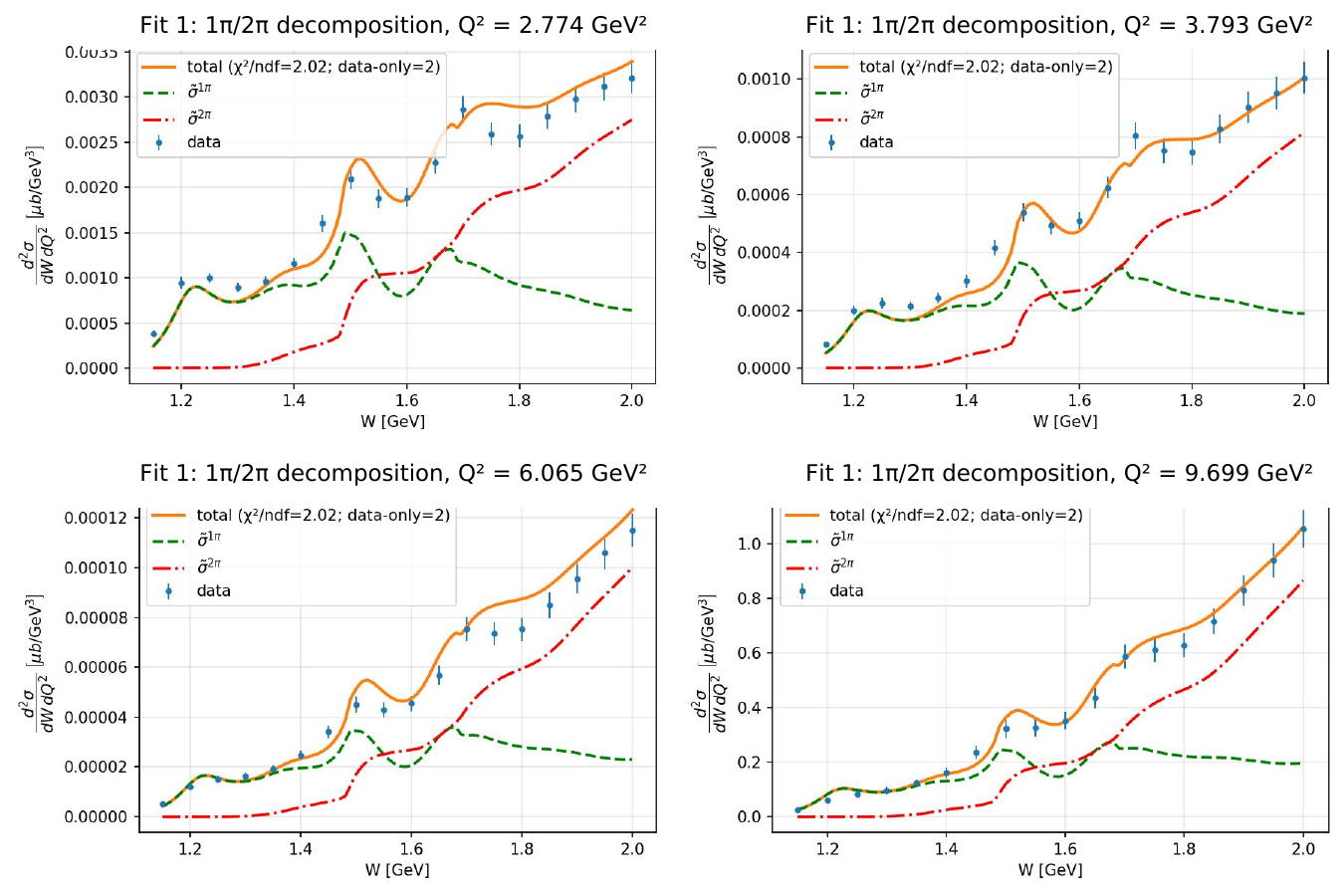}
\caption{Fit~1 decomposition of the CLAS12 inclusive cross section
$d^{2}\sigma/dW\,dQ^{2}$ into single-pion and multi-pion contributions,
shown in $W$ slices for four representative $Q^{2}$ bins,
$Q^{2} = 2.774$, $3.793$, $6.065$, and $9.699~\mathrm{GeV}^{2}$.
Blue points: CLAS12 RGA data~\cite{KlimenkoCLAS12}. Orange solid:
total Fit~1 result (data-only $\chi^{2}/\mathrm{ndf} \approx 2$), the
sum $\tilde\sigma^{1\pi}+\tilde\sigma^{2\pi}$ of
Eq.~(\ref{eq:fit1}). Green dashed: evolved
single-pion contribution $\tilde\sigma^{1\pi}$. Red dot-dashed: evolved
multi-pion contribution $\tilde\sigma^{2\pi}$. The $1\pi$ contribution
dominates the low-$W$ region, including the $\Delta(1232)$ peak, while
the $2\pi$ contribution carries the bulk of the high-$W$ strength.}
\label{fig:global_LT_1pi2pi}
\end{figure}

Figure~\ref{fig:global_LT_1pi2pi} shows the result of this fit at four
representative $Q^{2}$ bins. The total (orange solid) is the sum,
Eq.~(\ref{eq:fit1}), of the evolved single-pion contribution
$\tilde\sigma^{1\pi}$ (green dashed) and the evolved multi-pion
contribution $\tilde\sigma^{2\pi}$ (red dot-dashed). The $1\pi$
contribution governs the low-$W$ behavior, in particular the
$\Delta(1232)$ peak, whereas the $2\pi$ contribution grows in relative
importance with $W$ and carries the bulk of the high-$W$ strength at
every $Q^{2}$. Any $Q^{2}$ evolution of the relative $1\pi$/$2\pi$
balance is mild and parametrization-dependent within the present
inclusive data, and remains to be tested directly against forthcoming
CLAS12 exclusive $p(e,e'\pi)N$ and $p(e,e'\pi\pi)N$ measurements. This
stage fixes the anchored $Q^{2}$-evolution of the inclusive cross
section and its single-pion/multi-pion content over the resonance
region, $W \lesssim 2.0~\mathrm{GeV}$. The longitudinal--transverse
separation and the extension to $W \le 2.5~\mathrm{GeV}$ are carried out
in Fit~2 (Sec.~\ref{sec:fit2}).

\subsection{Fit~2: longitudinal--transverse decomposition
($R_{LT}$ constrained at $Q_{0}^{2}$)}
\label{sec:fit2}

The second stage decomposes the inclusive cross section into
longitudinal and transverse virtual-photon components, over the full
measured range $W \le 2.5~\mathrm{GeV}$,
\begin{align}
\sigma^{\rm tot}_{\rm AO}(Q^{2},W,E_{e})
\;=\;&
%\bm{c}_{\rm units}
\bigl[
\tilde{\sigma}^{\rm AO}_{T}(Q^{2},W)
+ \epsilon(Q^{2},W,E_{e})\,
\tilde{\sigma}^{\rm AO}_{L}(Q^{2},W)
\bigr]
\notag\\
&\times \Gamma_{\rm Hand}(Q^{2},W,E_{e})\,
\mathcal{J}(Q^{2},W,E_{e}),
\label{eq:sigma_LT}
\end{align}
where $\Gamma_{\rm Hand}$ is the Hand-convention virtual-photon flux,
$\epsilon$ is the transverse photon polarization, $\mathcal{J}$ is the
Jacobian to the $(W,Q^{2})$ phase space, and $E_{e}$ is the electron beam
energy.
%and $\bm{c}_{\rm units} = 389.379$ is a unit-conversionfactor. 
The kinematic quantities are defined by
\begin{align}
\epsilon(Q^{2},W,E_{e})
&= \left[
1 + 2\left(1 + \frac{\nu^{2}}{Q^{2}}\right)
\tan^{2}\frac{\theta_{e}}{2}
\right]^{-1},
\label{eq:epsilon}\\[4pt]
\Gamma_{\rm Hand}(Q^{2},W,E_{e})
&= \frac{\alpha_{\rm em}}{2\pi^{2}}\,
\frac{E_{e}-\nu}{E_{e}}\,
\frac{K(W)}{Q^{2}\,[1-\epsilon]},
\label{eq:gamma_hand}\\[4pt]
K(W) &= \frac{W^{2}-m_{N}^{2}}{2 m_{N}},
\qquad
\nu(Q^{2},W) = \frac{W^{2}+Q^{2}-m_{N}^{2}}{2 m_{N}},
\label{eq:KandNu}\\[4pt]
\mathcal{J}(Q^{2},W,E_{e})
&= \frac{\pi W}{m_{N} E_{e}\,(E_{e}-\nu)},
\label{eq:jacobian}\\[4pt]
\theta_{e}(Q^{2},W,E_{e})
&= \arccos\!\left[
1 - \frac{Q^{2}}{2 E_{e}\,(E_{e}-\nu)}
\right],
\label{eq:theta_e}
\end{align}
where $\alpha_{\rm em}$ is the fine-structure constant and $m_{N}$ is
the nucleon mass. The extended ANL-Osaka components are written in the
same anchored form as in Fit~1,
\begin{align}
\tilde{\sigma}^{\rm AO}_{T}(Q^{2},W)
&= f(Q^{2};\alpha)\,
\frac{F_{2}(\omega'[Q^{2},W];c_{1},c_{2},c_{3},c_{4})}
     {F_{2}(\omega'[Q_{0}^{2},W];c_{1},c_{2},c_{3},c_{4})}
\,\sigma^{\rm AO}_{T,{\rm input}}(Q_{0}^{2},W),
\label{eq:sigmaT}\\[4pt]
\tilde{\sigma}^{\rm AO}_{L}(Q^{2},W)
&= f(Q^{2};\beta)\,
\frac{F_{2}(\omega'[Q^{2},W];c'_{1},c'_{2},c'_{3},c'_{4})}
     {F_{2}(\omega'[Q_{0}^{2},W];c'_{1},c'_{2},c'_{3},c'_{4})}
\,\sigma^{\rm AO}_{L,{\rm input}}(Q_{0}^{2},W),
\label{eq:sigmaL}
\end{align}
with the ANL-Osaka input cross sections
$\sigma^{\rm AO}_{T,{\rm input}}$ and $\sigma^{\rm AO}_{L,{\rm input}}$
extended in $W$ up to $2.5~\mathrm{GeV}$ and the exponential scaling
functions
\begin{equation}
f(Q^{2};\alpha) = \exp[\alpha(Q^{2}-Q_{0}^{2})],
\qquad
f(Q^{2};\beta) = \exp[\beta(Q^{2}-Q_{0}^{2})].
\label{eq:expscaling}
\end{equation}
In this construction, the $W$ dependence of
$\sigma^{\rm AO}_{T,{\rm input}}(Q_{0}^{2},W)$ and
$\sigma^{\rm AO}_{L,{\rm input}}(Q_{0}^{2},W)$ is taken from the
original ANL-Osaka calculation and held fixed, while the parameters
$\alpha$, $\beta$, and $\{c_{i},c'_{i}\}$ control the phenomenological
$Q^{2}$ evolution constrained by the inclusive CLAS12 cross sections
and the soft $R_{LT}$ penalty. The inputs
$\sigma^{\rm AO}_{T,{\rm input}}$ and $\sigma^{\rm AO}_{L,{\rm input}}$
are the channel-summed transverse and longitudinal cross sections, so
Fit~2 resolves the longitudinal and transverse components of the total
inclusive cross section but does not separate the single-pion and
multi-pion contributions; that decomposition is provided only by Fit~1.

The modified Breidenbach--Kuti functions used here are extended,
relative to the three-term form of Eq.~(\ref{eq:F2_BK_3}) used in Fit~1,
by an additional $(1-1/\omega')^{6}$ term,
\begin{align}
F_{2}(\omega';c_{1},c_{2},c_{3},c_{4})
&= \sum_{i=1}^{4} c_{i}
\left(1 - \frac{1}{\omega'}\right)^{2+i},
\label{eq:F2_BK_4a}\\
F_{2}(\omega';c'_{1},c'_{2},c'_{3},c'_{4})
&= \sum_{i=1}^{4} c'_{i}
\left(1 - \frac{1}{\omega'}\right)^{2+i},
\label{eq:F2_BK_4b}
\end{align}
in order to capture the change in curvature observed in the high-$W$
data ($W \gtrsim 2.0~\mathrm{GeV}$) as $Q^{2}$ increases.

The parameters $\{\alpha,\beta,c_{i},c'_{i}\}$ are determined by
minimizing the total $\chi^{2}$,
\begin{equation}
\chi^{2}_{\rm total}
\;=\;
\chi^{2}_{\rm data}
\;+\;
\Phi_{\rm penalty},
\label{eq:chi2_total}
\end{equation}
where the data term is
\begin{equation}
\chi^{2}_{\rm data}
\;=\;
\sum_{\{Q_{i}^{2},W_{j}\}}
\frac{\bigl[
\sigma^{\rm tot}_{\rm data}(Q_{i}^{2},W_{j})
- \sigma^{\rm tot}_{\rm AO}(Q_{i}^{2},W_{j})
\bigr]^{2}}
{\Delta\sigma^{\rm tot}_{\rm data}(Q_{i}^{2},W_{j})^{2}},
\label{eq:chi2_data}
\end{equation}
and the penalty term implements a soft constraint on the
longitudinal-to-transverse ratio,
\begin{equation}
R_{LT}^{\rm AO,par.}(Q^{2},W)
\;\equiv\;
\frac{\tilde{\sigma}^{\rm AO}_{L}(Q^{2},W)}
     {\tilde{\sigma}^{\rm AO}_{T}(Q^{2},W)},
\label{eq:RLT}
\end{equation}
relative to the $R_{LT}$ values predicted by a reference model:
\begin{equation}
\Phi_{\rm penalty}
\;=\;
\sum_{\{Q_{{\rm m},i}^{2},W_{{\rm m},j}\}}
\frac{\bigl[
R_{LT}^{\rm model}(Q_{{\rm m},i}^{2},W_{{\rm m},j})
- R_{LT}^{\rm AO,par.}(Q_{{\rm m},i}^{2},W_{{\rm m},j})
\bigr]^{2}}
{\bigl[0.3\,
R_{LT}^{\rm model}(Q_{{\rm m},i}^{2},W_{{\rm m},j})
\bigr]^{2}}.
\label{eq:penalty}
\end{equation}

A soft (quadratic) penalty, rather than a hard constraint, is adopted so
that the data drive the fit where they are sufficiently sensitive, while
the result remains stabilized in regions of weaker experimental
constraint. The reference $R_{LT}$ values are obtained from the
ANL-Osaka model at $Q_{0}^{2}$ and extended as a $Q^{2}$-independent
function across the fitted range. A $30\%$ tolerance is assigned to the
reference $R_{LT}$ values in the penalty term, allowing substantial
deviations from the reference model while still providing sufficient
stabilization in regions where the inclusive cross sections alone have
limited sensitivity to $R_{LT}$.

The total $\chi^{2}$ surface is nonlinear in the fit parameters and
admits multiple local minima. To reliably identify the global minimum,
we employ a two-stage Monte Carlo search: $N_{\rm sampling} = 5000$
randomized initial parameter configurations are generated within
physically reasonable ranges
($\alpha,\beta \in [-4.5,0.0]$, $c_{i},c'_{i} \in [-600,600]$), the
$N_{\rm simplex} = 2000$ configurations with the smallest
$\chi^{2}_{\rm total}$ are refined by simplex
minimization~\cite{James1975}, and the best $N_{\rm MIGRAD} = 10$ of
these are subjected to gradient-based {\sc Migrad} minimization. The
global best-fit solution is selected as the configuration yielding the
smallest $\chi^{2}_{\rm total}$. This procedure was found essential to
avoid convergence to shallow local minima, particularly when alternative $R_{LT}$ priors are employed.

Because the Breidenbach--Kuti functions are linear in their parameters,
the overall normalizations $c_{1}$ and $c'_{1}$ cancel in the ratios of
Eqs.~(\ref{eq:sigmaT})--(\ref{eq:sigmaL}), so the model is fully
specified by the irreducible parameter set
\begin{equation}
\left\{
\alpha,\beta,\,
r_{1} = c_{2}/c_{1},\,
r_{2} = c_{3}/c_{1},\,
r_{3} = c_{4}/c_{1},\,
r'_{1} = c'_{2}/c'_{1},\,
r'_{2} = c'_{3}/c'_{1},\,
r'_{3} = c'_{4}/c'_{1}
\right\}.
\label{eq:irreducible}
\end{equation}
In practice the global minimum search is carried out in the
$\{c_{i},c'_{i}\}$ representation, which yields a smoother
$\chi^{2}$ landscape and improved numerical convergence in the Monte
Carlo procedure; the result is transformed to the irreducible
$\{r_{i},r'_{i}\}$ representation after the fact for the presentation of
final results and for the uncertainty estimation.

Statistical uncertainties on the fitted parameters were determined using
a parametric bootstrap procedure~\cite{Efron1979,EfronTibshirani1994}. A
total of $2000$ bootstrap replicas were generated by fluctuating the
measured cross sections according to their total experimental
uncertainties, obtained by combining statistical and systematic
contributions in quadrature and assuming Gaussian probability
distributions. Each bootstrap replica was analyzed using the full
constrained fitting procedure, including the $R_{LT}$ penalty term,
starting from the nominal best-fit solution. Because the entire nonlinear
fitting procedure is repeated for every replica, correlations among all
fit parameters are naturally incorporated into the resulting uncertainty
estimates. The quoted parameter uncertainties correspond to the standard
deviations of the bootstrap distributions and therefore represent the
propagation of experimental uncertainties through the complete fitting
procedure. This approach avoids reliance on the local quadratic
approximation of the $\chi^{2}$ surface, which does not provide reliable
uncertainty estimates for the highly nonlinear constrained model
considered here.

The uncertainties quoted in Table~\ref{tab:fit_params} represent the
experimental uncertainties propagated through the bootstrap procedure,
including the statistical and systematic uncertainties of the measured
cross sections added in quadrature. Additional model-dependent
systematic effects associated with the phenomenological extension of the
ANL-Osaka model have not yet been quantified separately. These include
possible sensitivity to the functional form of the high-$Q^{2}$
extrapolation, the order of the modified Breidenbach--Kuti
parametrization, the choice of fit range, and the imposed $R_{LT}$
constraint. Nevertheless, the extracted inclusive observables were found
to remain comparatively stable under the alternative $R_{LT}$ priors
considered in Table~\ref{tab:fit_params}. Using
the CJ15-constrained fit in place of the ANL-Osaka-constrained default
leads to only modest shifts of the extracted $F_{2}$ and $M_{2}$ values
in the resonance region, well within the total uncertainties quoted in
this work; the variation between the two constraints is smaller than the
experimental uncertainty over most of the kinematic range. A more
complete assessment of model-dependent systematic
uncertainties, including propagation into the extracted $F_{2}$ structure
functions and truncated moments, will be pursued in future work.

\begin{table}[htbp]
\centering
\caption{Best-fit parameters obtained for two different $R_{LT}$
constraints (total experimental uncertainties from the parametric
bootstrap procedure). $r_{i} \equiv c_{i+1}/c_{1}$ and
$r'_{i} \equiv c'_{i+1}/c'_{1}$ are the irreducible parameter
combinations that fully specify the model. The
$\chi^{2}_{\rm total}/\mathrm{ndf}$,
$\chi^{2}_{\rm data}/\mathrm{ndf}$, and
$\Phi_{\rm penalty}/\mathrm{ndf}$ are also quoted, with
$\mathrm{ndf} = N_{\rm data} - N_{\rm parameters}$. In the
column headers, AO denotes the ANL-Osaka constraint.}
\label{tab:fit_params}
\begin{ruledtabular}
\begin{tabular}{lcc}
Parameter & AO-constrained (adopted) & CJ15-constrained \\
\hline
$\alpha$ & $-0.069 \pm 0.011$ & $-0.119 \pm 0.009$ \\
$\beta$ & $-0.063 \pm 0.011$ & $-0.379 \pm 0.030$ \\
$r_{1}$ & $-1.758 \pm 0.499$ & $-2.508 \pm 0.252$ \\
$r_{2}$ & $\phantom{-}4.495 \pm 0.852$
        & $\phantom{-}3.858 \pm 0.537$ \\
$r_{3}$ & $-3.207 \pm 0.650$ & $-2.347 \pm 0.380$ \\
$r'_{1}$ & $\phantom{-}4.861 \pm 0.600$
         & $-13.726 \pm 0.127$ \\
$r'_{2}$ & $-16.958 \pm 1.670$
         & $\phantom{-}46.761 \pm 1.150$ \\
$r'_{3}$ & $\phantom{-}18.998 \pm 1.800$
         & $-52.834 \pm 2.480$ \\
\hline
$\chi^{2}_{\rm total}/\mathrm{ndf}$ & 2.02 & 2.58 \\
$\chi^{2}_{\rm data}/\mathrm{ndf}$  & 2.00 & 1.95 \\
$\Phi_{\rm penalty}/\mathrm{ndf}$   & 0.02 & 0.63 \\
\end{tabular}
\end{ruledtabular}
\end{table}

The best-fit parameters obtained with two alternative $R_{LT}$
constraints (ANL-Osaka and CJ15) are reported in
Table~\ref{tab:fit_params}. The data-only
$\chi^{2}_{\rm data}/\mathrm{ndf}$ is close to $2$ in both cases,
indicating that the experimental cross sections are well described
regardless of which $R_{LT}$ prior is used; the small difference in
$\chi^{2}_{\rm total}/\mathrm{ndf}$ reflects the different penalty
contributions. The nearly identical values of
$\chi^{2}_{\rm data}/\mathrm{ndf}$ obtained with the ANL-Osaka and CJ15
constraints further indicate that the inclusive cross sections
themselves do not strongly favor either $R_{LT}$ prior. The
ANL-Osaka-constrained fit is adopted as the default in what follows. The
stability of the extracted observables under these substantially
different $R_{LT}$ constraints indicates that the principal physics
conclusions are determined primarily by the CLAS12 data rather than by
the specific choice of phenomenological prior.

The extended ANL-Osaka model introduced here is not a substitute for a
full coupled-channels re-determination of the resonance form factors at
high $Q^{2}$; it provides a phenomenologically constrained
interpolation, anchored to the data and to the ANL-Osaka description at
$Q_{0}^{2}$, sufficient for the L/T decomposition and the duality
analysis presented below. No new resonance states are introduced, and no
assumptions are made regarding modifications of the underlying hadronic
interaction dynamics.

% =====================================================================
\section{Inclusive cross sections at multiple $Q^{2}$ bins}
\label{sec:xsec_4bins}
% =====================================================================

\begin{figure}[htbp]
\centering
\includegraphics[width=0.95\textwidth]{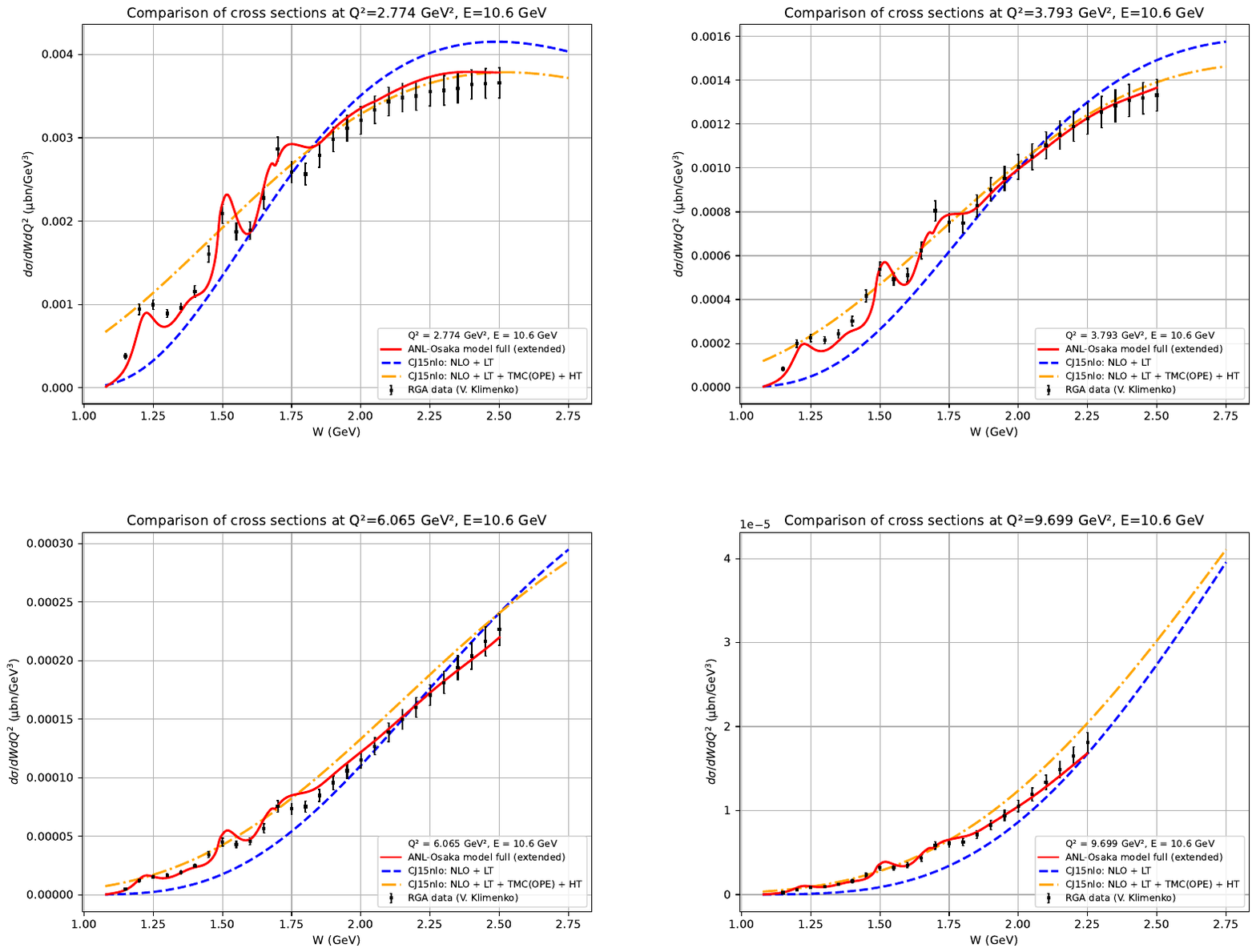}
\caption{Inclusive cross sections $d^{2}\sigma/dW\,dQ^{2}$ at
$Q^{2} = 2.774$, $3.793$, $6.065$, and
$9.699~\mathrm{GeV}^{2}$, shown over the full measured range up to
$W \approx 2.5~\mathrm{GeV}$. Black points: CLAS12 RGA
data~\cite{KlimenkoCLAS12}. Red solid: extended ANL-Osaka
calculation (the Fit~2 result of Sec.~\ref{sec:fit2}), anchored to the
original ANL-Osaka model at $Q_{0}^{2} = 2.774~\mathrm{GeV}^{2}$. Blue
dashed: CJ15NLO+LT. Orange dot-dashed: CJ15NLO+TMC+HT. The extended
ANL-Osaka calculation reproduces the observed resonance structure across
the full $Q^{2}$ range, while the inclusion of target-mass and
higher-twist corrections substantially improves agreement between the
CJ15 calculation and the data at larger $W$.}
\label{fig:xsec_4panel}
\end{figure}

Figure~\ref{fig:xsec_4panel} shows the measured CLAS12 inclusive cross
sections $d^{2}\sigma/dW\,dQ^{2}$ at four representative $Q^{2}$ values
($Q^{2} = 2.774$, $3.793$, $6.065$, and $9.699~\mathrm{GeV}^{2}$), over
the full measured range up to $W \approx 2.5~\mathrm{GeV}$, compared with
the extended ANL-Osaka calculation (red solid; the Fit~2 result of
Sec.~\ref{sec:fit2}), the CJ15NLO+LT prediction (blue dashed), and the
CJ15NLO+TMC+HT calculation (orange dot-dashed).

Several features are evident. First, the clear three-resonance
structure, dominant at $Q^{2} = 2.774~\mathrm{GeV}^{2}$, persists across
the full $Q^{2}$ range probed and is reproduced by the extended
ANL-Osaka calculation at every $Q^{2}$ point. Second, the CJ15NLO+LT
curve systematically undershoots the data at low $W$ and merges with the
data only at the largest $W$ and $Q^{2}$. Third, the inclusion of
target-mass and higher-twist corrections restores agreement with the
data above $W \approx 1.7~\mathrm{GeV}$ at all $Q^{2}$. Finally, the extended ANL-Osaka calculation describes both the
resonance structure and the average cross-section strength across the
resonance region, providing a continuous hadronic description that
merges smoothly with the CJ15NLO+TMC+HT prediction at higher $W$.

A qualitative difference between the hadronic and partonic descriptions
is visible near the pion-production threshold,
$W_{\rm thr} = m_{N} + m_{\pi} \approx 1.08~\mathrm{GeV}$. The inclusive
inelastic cross section must vanish at $W_{\rm thr}$, since no hadronic
final state is kinematically accessible below it. The extended
ANL-Osaka calculation satisfies this boundary condition by construction,
because it is built from explicit meson--baryon channels, each carrying
the correct threshold behavior. The CJ15NLO+LT calculation also vanishes
at $W_{\rm thr}$, as is evident from the blue dashed curve in
Fig.~\ref{fig:xsec_4panel}: the leading-twist cross section is driven by
the parton distributions evaluated at the corresponding Bjorken $x$, and
the relevant phase-space factor drives it smoothly to zero at the
threshold. When target-mass and higher-twist corrections are added,
however, this property is lost. The target-mass corrections shift the
argument of the structure functions from $x$ to the Nachtmann variable
$\xi$, and the phenomenological higher-twist term is a smooth analytic
function fitted to the inclusive data at larger $W$; neither contains
information about the physical $\pi N$ threshold. As a result, the
CJ15NLO+TMC+HT curve (orange dot-dashed) does not vanish at $W_{\rm thr}$.
Because the structure function is a smooth function of $W$, this
incorrect boundary value is not confined to the immediate vicinity of
$W_{\rm thr}$: starting from a nonzero value at threshold, the
CJ15NLO+TMC+HT curve also follows a different slope through the low-$W$
region, so that the distortion of its shape extends across the
first-resonance region and into the adjacent range of higher $W$. This
behavior is not a deficiency of the global QCD analysis; it simply
reflects the fact that the finite-$Q^{2}$ corrections, designed to
extend a leading-twist description to lower $W$ in an averaged sense,
are not expected to hold pointwise at the lowest invariant masses, where
the cross section is governed by exclusive hadronic dynamics. As
discussed in Secs.~\ref{sec:F2} and \ref{sec:moments}, this low-$W$
shape distortion propagates into the extracted $F_{2}$ and, in an
integrated form, into the first-resonance-region truncated moment. 

% =====================================================================
\section{Extraction of $F_{2}$ and local quark--hadron duality}
\label{sec:F2}
% =====================================================================

Using the L/T decomposition enabled by the extended ANL-Osaka
description, we determine the proton structure function
$F_{2}(W,Q^{2})$ from the measured inclusive cross sections. The
separation into longitudinal and transverse components is
phenomenological, constrained by the $R_{LT}$ penalty of
Sec.~\ref{sec:fit2}; the $F_{2}$ values reported here are therefore
obtained within this AO-constrained framework rather than from a
model-independent L/T separation of the data.

\begin{figure}[htbp]
\centering
\includegraphics[width=0.80\columnwidth]{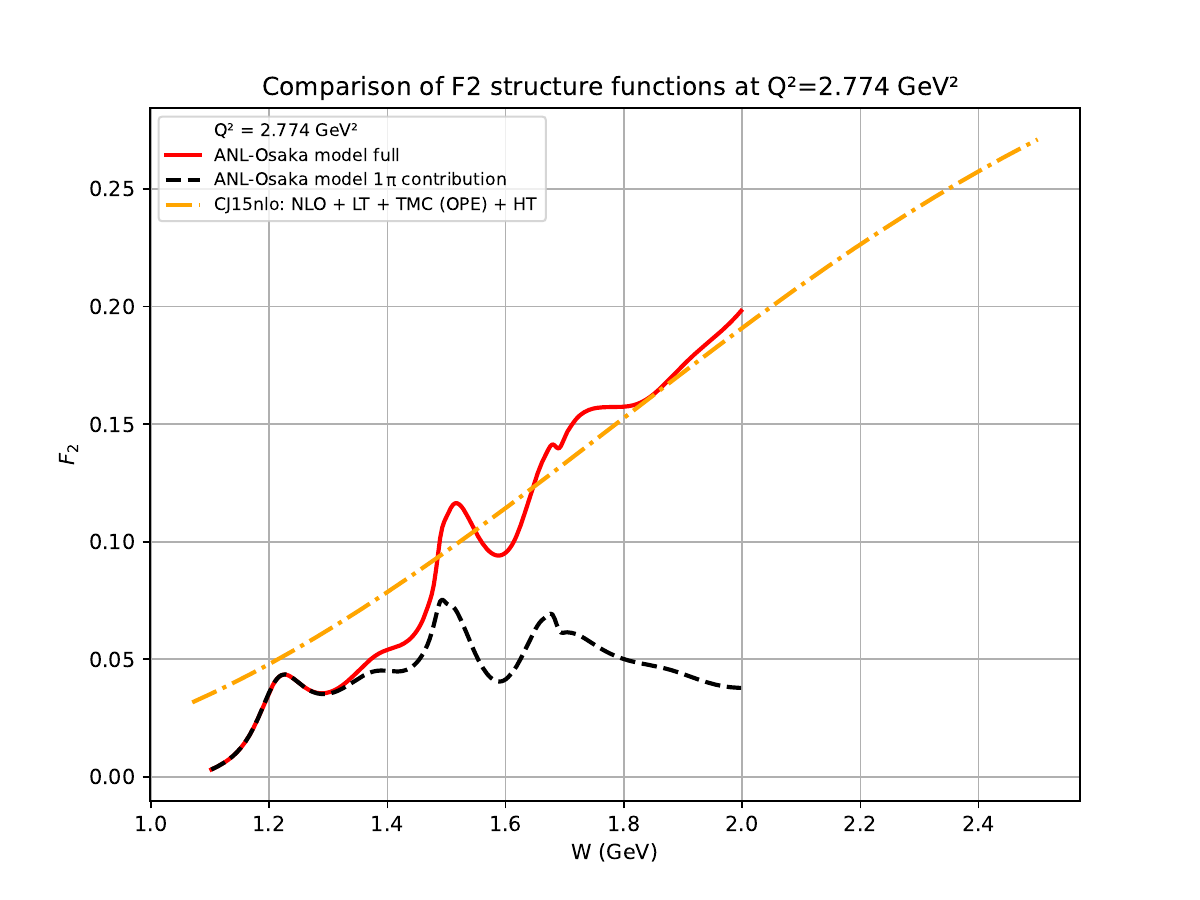}
\caption{Proton structure function $F_{2}$ at
$Q^{2} = 2.774~\mathrm{GeV}^{2}$. Red solid: full extended
ANL-Osaka calculation. Black dashed: $1\pi$ subchannel of the
ANL-Osaka model. Orange dot-dashed: CJ15NLO+TMC+HT. The full
ANL-Osaka calculation oscillates around the smooth CJ15NLO+TMC+HT
curve, illustrating local quark--hadron duality; the $1\pi$ piece
alone lies systematically below the smooth curve above the
$\Delta(1232)$ region, demonstrating the importance of multi-meson
channels in recovering the inclusive strength required by local
duality.}
\label{fig:F2_2p77_AO_vs_CJ15}
\end{figure}

Figure~\ref{fig:F2_2p77_AO_vs_CJ15} compares, at
$Q^{2} = 2.774~\mathrm{GeV}^{2}$, the full extended ANL-Osaka
prediction for $F_{2}$ (red solid) with its $1\pi$ subchannel (black
dashed) and with the CJ15NLO+TMC+HT calculation (orange
dot-dashed). The comparison illustrates the duality picture directly:
the full hadronic ANL-Osaka description produces the resonance peaks
and dips displayed by the data. The crossings between the
ANL-Osaka and CJ15NLO+TMC+HT curves occur near the dips between
resonances, providing a visual manifestation of local duality: the resonance-region
structure function oscillates around the smooth perturbative curve
and, when averaged, follows it. By contrast, the $1\pi$ subchannel
alone falls systematically below the CJ15NLO+TMC+HT curve above the
$\Delta(1232)$ region; at the structure-function level, the smooth
partonic strength is recovered only once the multi-meson channels are
included.

\begin{figure}[htbp]
\centering
\includegraphics[width=0.95\textwidth]{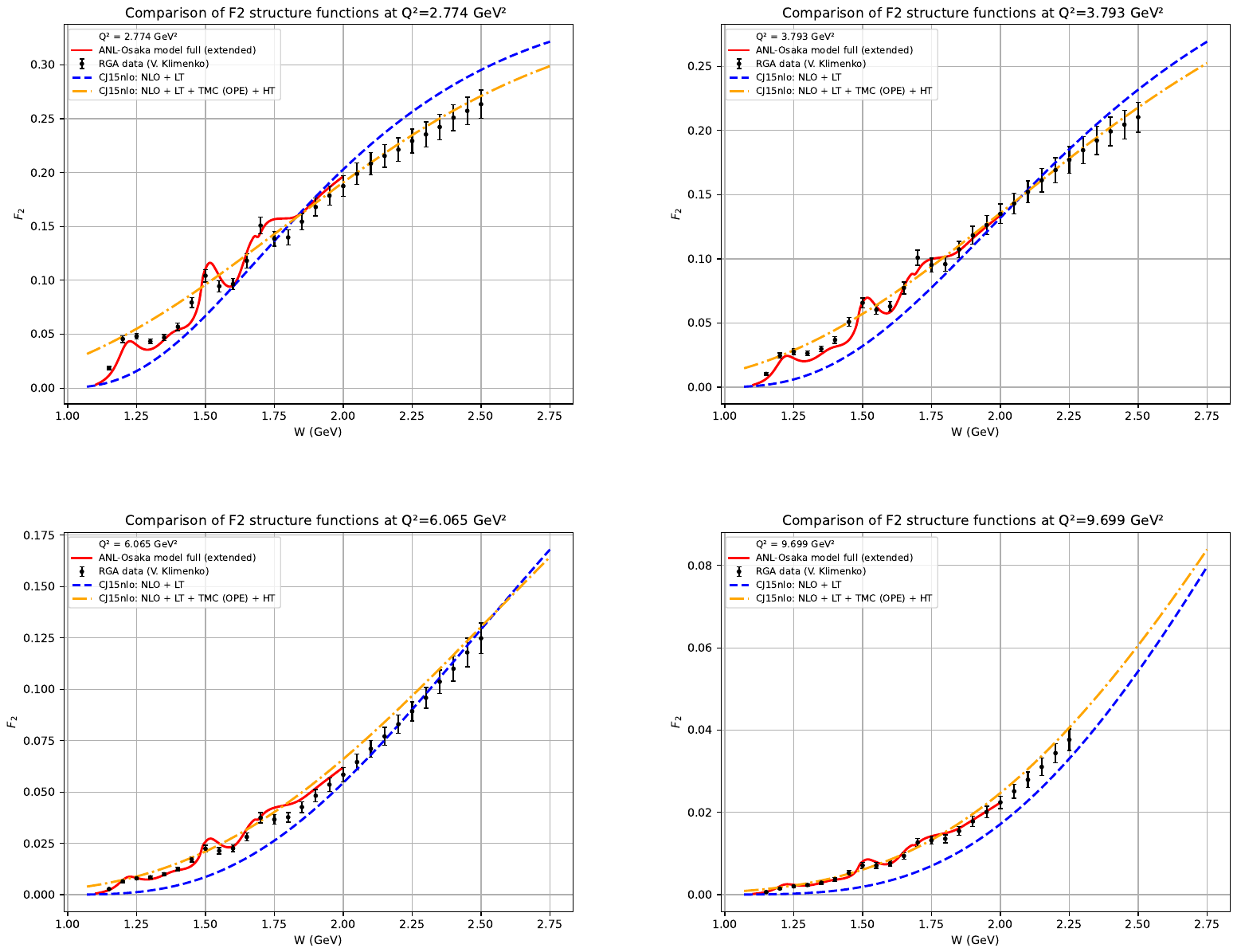}
\caption{Extracted proton structure function $F_{2}(W,Q^{2})$ at
$Q^{2} = 2.774$, $3.793$, $6.065$, and
$9.699~\mathrm{GeV}^{2}$, obtained after the L/T decomposition
using the extended ANL-Osaka description. Curves are as in
Fig.~\ref{fig:xsec_4panel}. The CJ15NLO+LT calculation (blue dashed) systematically undershoots the
extracted $F_{2}$ at low $W$, whereas the CJ15NLO+TMC+HT calculation
(orange dot-dashed) follows the average resonance-region strength. The oscillation of the extracted $F_{2}$
about the smooth CJ15NLO+TMC+HT curve at all measured $Q^{2}$ is
consistent with local quark--hadron duality.}
\label{fig:F2_4panel}
\end{figure}

Figure~\ref{fig:F2_4panel} shows the extracted $F_{2}(W,Q^{2})$ at the
same four representative $Q^{2}$ values, compared with the
CJ15NLO+LT calculation (blue dashed), CJ15NLO+TMC+HT
(orange dot-dashed), and the extended ANL-Osaka model (red
solid). The local-duality picture established at the lowest $Q^{2}$
extends across the full range: at every $Q^{2}$, the extracted $F_{2}$
retains the resonance structure seen in the data and oscillates around
the smooth CJ15NLO+TMC+HT curve, consistent with local quark--hadron
duality within the present uncertainties. 
The CJ15NLO+LT curve alone
systematically undershoots the extracted $F_{2}$ at low $W$, with the gap closing as
$Q^{2}$ increases, consistent with the expected suppression of
higher-twist contributions at large $Q^{2}$ in the operator product
expansion. 

Although the overall duality picture is robust, a residual difference
between the hadronic and partonic descriptions remains near the
pion-production threshold. Consistent with the cross-section
comparison of Sec.~\ref{sec:xsec_4bins}, the extracted $F_2$ and the CJ15NLO+LT curve vanish at $W_{\rm thr}$, whereas the CJ15NLO+TMC+HT curve does not, leading to a distortion of the low-$W$ behavior.

% =====================================================================
\section{Truncated moments and the onset of global duality}
\label{sec:moments}
% =====================================================================

A quantitative test of quark--hadron duality is provided by the
$W$-truncated Cornwall--Norton moments~\cite{CornwallNorton1969} of the
$F_{2}$ structure function,
\begin{equation}
M_{2}^{\rm trunc}(Q^{2};W_{\min},W_{\max})
\;=\;
\int_{x(W_{\max})}^{x(W_{\min})}
F_{2}(x,Q^{2})\,dx,
\label{eq:M2_trunc}
\end{equation}
where $x = Q^{2}/(W^{2} - m_{N}^{2} + Q^{2})$ is the Bjorken scaling
variable. By restricting the integration to a finite $W$ range and
varying that range, one isolates the contribution of individual
resonance regions and tests how well the smooth perturbative
description reproduces the integrated resonance strength.

\begin{figure}[htbp]
\centering
\includegraphics[width=0.95\textwidth]{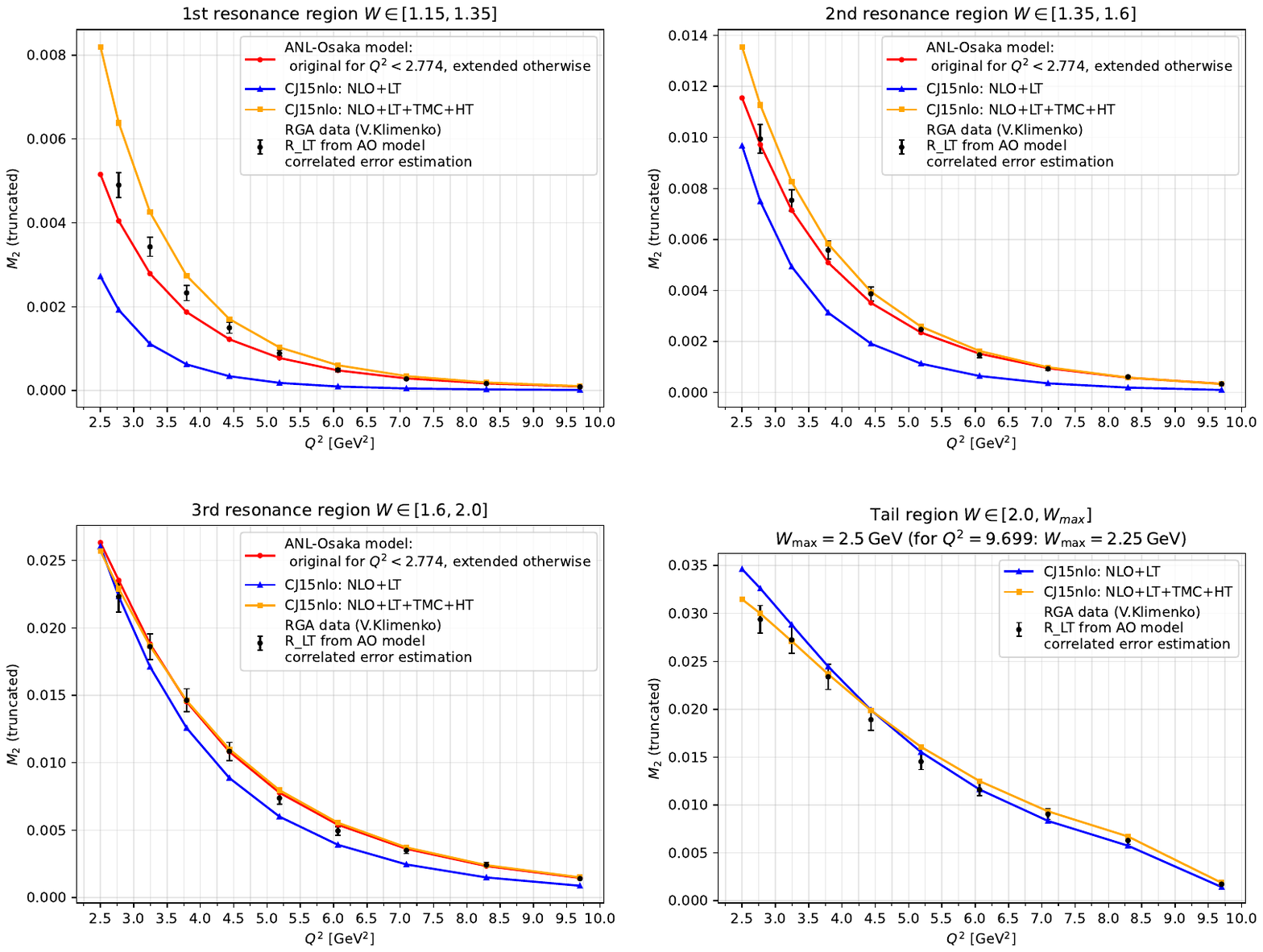}
\caption{Truncated $M_{2}$ moment of $F_{2}$ as a function of $Q^{2}$
for four $W$ ranges: the first resonance region (top left), the second
resonance region (top right), the third resonance region (bottom left),
and the high-$W$ region (bottom right). Black points: CLAS12 data with
correlated error estimates obtained using $R_{LT}$ from the extended
ANL-Osaka model. Red: extended ANL-Osaka model. Blue triangles:
CJ15NLO+LT. Orange squares: CJ15NLO+TMC+HT. Agreement between the extracted moments and the
CJ15NLO+TMC+HT calculation is observed across the second and third
resonance regions and the high-$W$ region, while the residual excess in
the first resonance region is consistent with the threshold effect
discussed in Secs.~\ref{sec:xsec_4bins} and \ref{sec:F2}.}
\label{fig:M2}
\end{figure}

Figure~\ref{fig:M2} shows the extracted $M_{2}^{\rm trunc}(Q^{2})$ for
four $W$ ranges: the first resonance region
$W \in [1.15,1.35]~\mathrm{GeV}$, the second resonance region
$W \in [1.35,1.60]~\mathrm{GeV}$, the third resonance region
$W \in [1.60,2.00]~\mathrm{GeV}$, and the high-$W$ region
$W \in [2.00,W_{\max}]$ with $W_{\max}=2.5~\mathrm{GeV}$ (reduced to
$2.25~\mathrm{GeV}$ at $Q^{2}=9.699~\mathrm{GeV}^{2}$ owing to
kinematic limits). The data are compared with the extended ANL-Osaka
calculation (red), CJ15NLO+LT (blue triangles), and CJ15NLO+TMC+HT
(orange squares). The principal observations are as follows.

In the first resonance region, $W \in [1.15,1.35]~\mathrm{GeV}$, the
CJ15NLO+LT curve lies well below the data over the entire
$Q^{2}$ range, reflecting the resonance strength carried by the
$\Delta(1232)$ that a smooth leading-twist description cannot reproduce.
The CJ15NLO+TMC+HT curve, by contrast, lies systematically above the
data, with the largest overshoot at the lowest $Q^{2}$, while the
extended ANL-Osaka calculation tracks the data closely. Moving to the
second resonance region, $W \in [1.35,1.60]~\mathrm{GeV}$, the
CJ15NLO+TMC+HT curve still exceeds the data at lower $Q^{2}$ values,
but the overshoot is visibly smaller than in the first region. In the
third resonance region, $W \in [1.60,2.00]~\mathrm{GeV}$, the
CJ15NLO+TMC+HT and ANL-Osaka curves agree closely with each other and
with the data across the full $Q^{2}$ range, while in the high-$W$
region $W \in [2.00,W_{\max}]$ the CJ15NLO+TMC+HT moment and the data
are likewise in good agreement. In all four $W$ ranges the
CJ15NLO+LT curve lies below the CJ15NLO+TMC+HT curve, and the gap
between the two narrows as $Q^{2}$ increases.

Two systematic trends are therefore visible in Fig.~\ref{fig:M2}.
First, in each individual resonance region the agreement between the
resonance-region data and the CJ15NLO+TMC+HT moment improves as $Q^{2}$
increases, which is the expected onset of quark--hadron duality as
higher-twist contributions are suppressed. Second, and independently of
$Q^{2}$, the CJ15NLO+TMC+HT moment shifts from a clear overshoot in the
first resonance region, to a reduced overshoot in the second, to close
agreement in the third region and the high-$W$ region. This ordering,
with the largest deviation nearest the pion-production threshold and
progressively smaller deviations at higher invariant mass, follows the
same pattern as the threshold boundary-condition mismatch discussed in
Sec.~\ref{sec:xsec_4bins}.

The low-$W$ shape distortion of the CJ15NLO+TMC+HT curve introduces an
additional contribution to the truncated moments that is largest in the
first resonance region and decreases progressively with increasing
invariant mass. The observed ordering of the deviations in
Fig.~\ref{fig:M2} is consistent with the threshold effect introduced in
Sec.~\ref{sec:xsec_4bins}. Within the quoted uncertainties, the second
and third resonance regions and the high-$W$ region are in good
agreement with CJ15NLO+TMC+HT for
$Q^{2}\gtrsim2.5~\mathrm{GeV}^{2}$, while the first resonance region
retains a modest excess that is consistent with this threshold effect.
A more quantitative assessment of threshold-constrained finite-$Q^{2}$
corrections lies beyond the scope of the present work. 

The four $W$ ranges from $W=1.15~\mathrm{GeV}$ to $W_{\max}$ are
contiguous and together span the entire resonance region and the
high-$W$ region beyond. This allows the relationship between local and
global duality to be examined directly. A moment integrated over the full resonance
region is, by the additivity of the integral in
Eq.~(\ref{eq:M2_trunc}), simply the sum of the individual
contributions. Because the four resonance intervals are contiguous,
agreement in each individual region implies agreement for their sum,
namely the moment integrated over the full resonance region. In this
sense, local duality in the truncated-moment analysis naturally leads to
global duality, whereas the converse does not necessarily hold, since a
global match could in principle arise from compensating deviations among
different resonance regions.

The present results establish agreement between the extracted moments
and the CJ15NLO+TMC+HT expectation across the second and third
resonance regions and the high-$W$ region, while only a modest excess
remains in the first resonance region near threshold. Since this excess is consistent with the threshold
boundary-condition effect discussed above rather than with a breakdown
of duality, the data provide strong evidence that global
quark--hadron duality is realized across the full
$Q^{2}$ range probed by CLAS12.

% =====================================================================
\section{Summary and outlook}
\label{sec:summary}
% =====================================================================

We have presented a high-precision study of quark--hadron duality in
inclusive electron--proton scattering extending to
$Q^{2}\approx10~\mathrm{GeV}^{2}$ across the full resonance region using recent CLAS12 measurements and an extended ANL-Osaka dynamical
coupled-channels framework.

The analysis establishes consistency simultaneously at the level of
inclusive cross sections, extracted structure functions, and truncated
Cornwall--Norton moments, providing new quantitative evidence for both
local and global quark--hadron duality at substantially higher momentum
transfers than previously explored. The extracted resonance-region
structure functions oscillate about the smooth perturbative expectation,
while the truncated moments demonstrate agreement region by region
across the resonance spectrum. Taken together, these observations show
that local duality is realized throughout the resonance region and,
through the additivity of the truncated moments, imply global duality
over the full resonance region. Comparison with the CJ15 global QCD analysis, including target-mass
corrections and phenomenological higher-twist contributions, indicates
that the residual discrepancy observed in the first resonance region is
consistent with the threshold boundary-condition artifact discussed
above. This threshold effect---that the finite-$Q^{2}$ corrections in
global PDF analyses do not enforce the physical pion-production
threshold---points to a concrete refinement of such analyses for future
duality tests.

Several directions for future work follow naturally from the present
analysis. The extended ANL-Osaka model can be used to compute higher
truncated
moments ($M_{4}$, $M_{6}$) and thereby constrain the higher-twist matrix
elements through the operator product expansion. A related direction
concerns the threshold boundary condition: as discussed in
Sec.~\ref{sec:moments}, the finite-$Q^{2}$ corrections applied to the
CJ15NLO+LT baseline do not preserve the vanishing of the structure function
at $W_{\rm thr}$, and a global-analysis prescription that enforces this
constraint would reduce the first-resonance-region moments and further
sharpen the duality comparison. The determined factors
$e^{\alpha(Q^{2}-Q_{0}^{2})}F_{2}(\omega';\{c_{i}\})$ and
$e^{\beta(Q^{2}-Q_{0}^{2})}F_{2}(\omega';\{c'_{i}\})$ can be used to
inform a high-$Q^{2}$ extension of the ANL-Osaka $\gamma N \to N^{*}$
form-factor parametrization,
Eq.~(\ref{eq:FN_poly}), in preparation for fits to
forthcoming CLAS12 exclusive $p(e,e'\pi)N$ and $p(e,e'\pi\pi)N$ data in
the range $Q^{2} = 3$--$10~\mathrm{GeV}^{2}$. The duality study can
also be extended to projected JLab~22~GeV
kinematics~\cite{Arrington2022_22GeV} ($E = 22~\mathrm{GeV}$, $Q^{2} =
12$--$18~\mathrm{GeV}^{2}$) using the extended ANL-Osaka framework, in
support
of the scientific case for the energy upgrade. Finally, the extracted $F_{2}$ values can be compared with, and
incorporated into, new global PDF analyses, to assess the impact of these
resonance-region measurements on the large-$x$ proton PDFs and to further
illuminate the transition between the hadronic and partonic descriptions
of nucleon structure.

\begin{acknowledgments}
This work was supported by the U.S. Department of Energy, Office of Science, under Contract Nos. DE-AC05-06OR23177, DE-AC02-05CH1123, and Contract No. DE-FG02-04ER41309.
\end{acknowledgments}

\bibliographystyle{apsrev4-2}
\bibliography{references}

\end{document}